# LINKING SAP FLOW MEASUREMENTS WITH EARTH OBSERVATIONS

*Enrico Tomelleri, Giustino Tonon*

Free University of Bozen-Bolzano

## ABSTRACT

While single-tree transpiration is challenging to compare with earth observation, canopy scale data are suitable for this purpose. To test the potentialities of the second approach, we equipped the trees at two measurement sites with sap flow sensors in spruce forests. The sites have contrasting topography. The measurement period covered the months between June 2020 and January 2021. To link plot scale transpiration with earth observations, we utilized Sentinel-2 and local meteorological data. Within a machine learning framework, we have tested the suitability of earth observations for modelling canopy transpiration. The $R^2$ of the cross-validated trained models at the measurement sites was between 0.57 and 0.80. These results demonstrate the relevance of Sentinel-2 data for the data-driven upscaling of ecosystem fluxes from plot scale sap flow data. If applied to a broader network of sites and climatic conditions, such an approach could offer unprecedented possibilities for investigating our forests' resilience and resistance capacity to an intensified hydrological cycle in the contest of a changing climate.

## 1. INTRODUCTION

Evapotranspiration (ET) is the sum of water released by evaporation and plant transpiration from a vegetated surface. The first component accounts for the water transfer from sources such as the soil and canopy interception to the air. The second component accounts for the water transfer occurring through vegetation. While the first is driven mainly by water availability and local meteorological conditions, the second pathway is strongly modulated by plant physiology. The regulation of the transpiration through the stomata is determined by a trade-off between increasing photosynthesis and contrasting desiccation. At the same time, transpiration is a crucial component of the leaf energy balance. Transpiration also provides the driving force for nutrient uptake and transport from roots to leaves. All of this makes transpiration a critical factor in the global carbon and water cycles. Therefore, it is crucial to investigate this process's ecosystem-scale dynamics to understand better vegetation feedbacks to an intensified hydrological cycle [1].

Notwithstanding their fundamental role in the functioning of the Earth-System, transpiration and its spatiotemporal patterns are not well constrained by currently available observations and poorly represented in ecological models [2] for predictions of the regional hydrological and carbon cycles under current and future climatic conditions.

Broad-scale initiatives like SAPFLOWNET [3] provide new opportunities for addressing this gap by making available harmonized datasets undergoing standardized quality assurance and quality control procedures

Still, scaling transpiration from trees to landscape is a fundamental problem in spatial ecohydrology. This gap can be addressed by integrating *in-situ* data with earth observations. Such integration can only happen if the *in-situ* and remote observations are conducted at a comparable spatial and temporal scale. In this context, we demonstrate with this study the effectiveness of a machine learning approach to combine Sentinel-2 data with weekly sap flow measurements at the plot scale.

## 2. METHODS

### 2.1. The study sites

The study sites (Fig. 1) are situated in the eastern Alps over Spruce forests and are part of the Italian tree-talkers network. Site-1 (Val d'Ultimo) is located at 1111 m a.s.l. (46.594°N, 11.077°E) and site-2 (Carezza) is located at 1653 m a.s.l. (46.404°N 11.588°E). While the precipitation regime is comparable, site-1 is situated in on steep terrain with a South-East aspect while site-2 on a relatively flat terrain with a North aspect.

We mapped and measured all the trees within a 12 m radius at each site. Finally, we equipped 20 trees per plot with IoT based Tree-Talkers (NATURE 4.0 SB Srl). The trees to be equipped have been selected randomly among those in the dominant plane. The measurement devices are suitable for measuring environmental variables as well tree specific parameters [4]. The devices are connected within a local LoRa network to a gateway. This collects the data and submits them to a web server via GPRS.

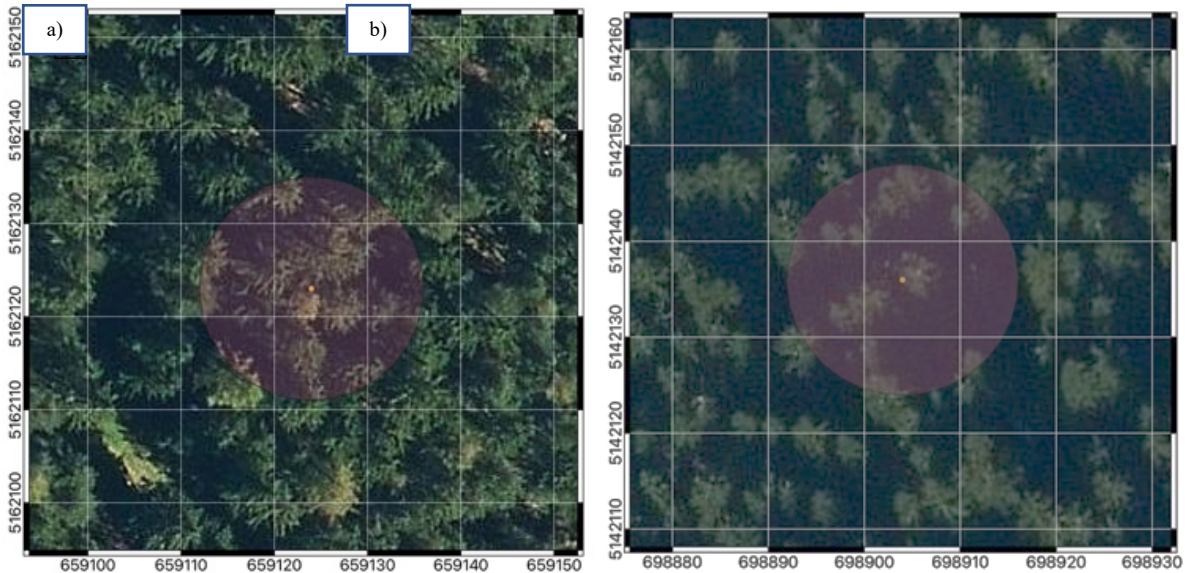

Figure 1: locations of test sites (coordinates in UTM zone 32N). a) Site-1 (Val d'Ultimo); b) Site-2 (Carezza). In the background an orthophoto with 0.2 m ground resolution (Source: Autonomous Province of Bozen/Bolzano, 2014/2015).

### 2.2. The Sap-Flow measurements

The Tree-Talkers are equipped with thermal dissipation probes [5]. Starting from June 2020, we acquired sap-flow measurement on an hourly basis. Raw-data have been acquired and pre-processed by means of the "ttalkR" R-Package [6]. For upscaling from tree to canopy, firstly, we aggregated the data to daily fluxes. Secondly, we estimated the tree specific sapwood area from the DBH measured during the initial survey as described by [7]. Lately, we weighted the sap flow density for each tree sap flow area to obtain the plot scale daily transpiration. We averaged daily transpiration to weekly time scale for the following analyses. This is necessary to match the temporal scale of *in-situ* measurements with the Sentinel-2 observations.

### 2.3. The remote sensing datasets

We obtained the datasets by sub-setting the Google Earth Engine collection ("COPERNICUS/S2_SR"). In doing this, we averaged the values of the pixels within a 12 m radius from the central coordinates at each study site. By doing this, we overcome potential issues related to possible geolocation errors for those bands with 10 m resolution when using the coordinates of a single pixel. The used data collection corresponds to Sentinel-2 L2 data as provided from sci-hub. Data were computed by running sen2cor. Initially, we filtered Sentinel-2 time series with the cloud and snow flags available in the original dataset. Then, we aggregated the time series to weekly values. We opted for this time scale because - given the overpass frequency of Sentinel-2 over the area of interest – the weekly detection of variations in vegetation optical properties is suitable providing continuous enough time series without incurring in big gaps caused by the atmospheric variability in mountainous regions.

### 2.4. The meteorological data

We made use of local weather stations to be able to analyze the whole time series of sap flow measurements. In fact, the most recent period is not covered by gridded high-resolution meteorological dataset like E-Obs. Therefore, we obtained daily temperature and precipitation for the specific period from the nearest weather station to each site. The one used for site-1 is located at 1142 m a.s.l (46.542 °N, 10.989 °E). The weather station for site-2 is located at 1128 m a.s.l (46.429 °N, 11.537 °E). As we did for the remote sensing data, we aggregated the meteorological variables to a weekly time scale.

### 2.5. The machine learning framework

For modelling canopy level transpiration with remote sensing and meteorological data, we selected the Random Forest (RF) algorithm [8]. For this purpose, we made use of the "randomForest" R-package [9]. For tunning the algorithm, it is necessary to identify the number of randomly collected variables to be sampled at each split time. For this purpose, we adopted a 5-fold, 30 repeat cross-validation where RMSE was used to select the optimal model. The metrics reported in Tab. 1 refer to the average of best performing model over the cross-validation procedure. In addition to assessing a model's overall predictive ability at the test sites with different combinations of predictive datasets, we quantified the contribution of each predictor variable to model performance by using the residual sum of squares according to [10].

### 3. RESULTS AND DISCUSSION

The weekly time series of plot scale transpiration (Fig. 2) show a decrease of fluxes from summer to winter with some peaks which are not synchronized in the two sites. In fact, during the autumn, we detect a peak of transpiration at the end of September at site-1 and mid-October at site-2. This might be because the local meteorological conditions are mediating the week-to-week variability while the regional weather situation drives the seasonality.

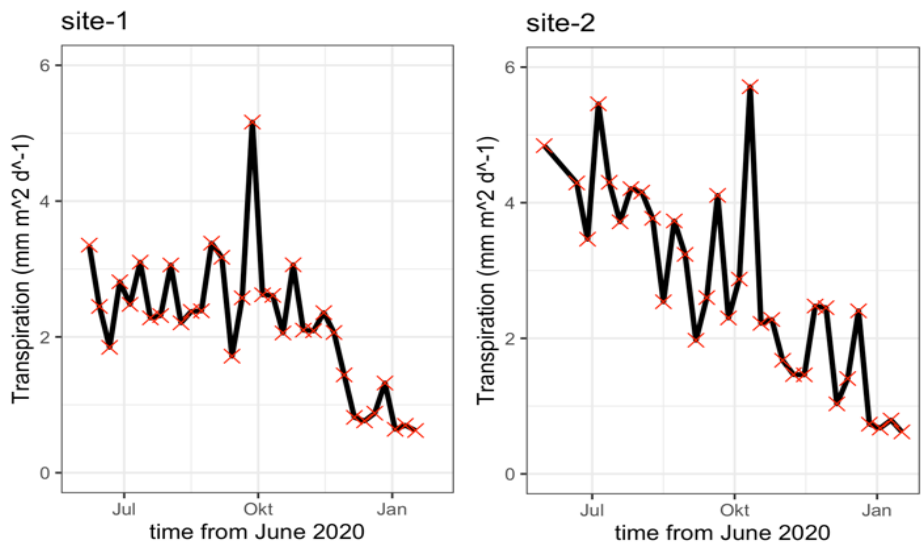

Figure 2: canopy scale transpiration at the measurement sites

The inclusion of meteorological data in the statistical models shows an improvement for site-1, where the $R^2$ was relatively low using only Sentinel-2 data (Tab. 1). On the other side, the relative improvement with meteorology at site-2 is reduced. In the first case, the steep terrain might affect the quality and the correct geolocation of remote sensing data. Additionally, while the weather station is located on flat terrain at the centre of the valley and near a lake, the measurement plot is on a slope exposed to the South-East. For this reason, the temperature variability might be buffered at the weather station and accentuated at the plot. This because of solar radiation.

Table 1: $R^2$ and mean absolute error of the trained models at the measurement sites

| Site | Sentinel-2 | Sentinel-2+Meteo |
|---|---|---|
| 1 | 0.28 (0.68) | 0.57 (0.50) |
| 2 | 0.79 (0.70) | 0.80 (0.68) |

The relative importance of variables (Tab. 2) shows that the short-wave infrared region (SWIR) is relevant for modelling transpiration with and without meteorological variables. On the other hand, there isn't a clear pattern across the sites and the experimental setups within the bands in the visible. Again, this might be associated with the challenges of remote sensing in complex terrains.

Table 2: Relative importance of variables

| Ranking | Site-1 S-2 | Site-2 S-2 | Site-1 S2+Meteo | Site-2 S2+Meteo |
|---|---|---|---|---|
| 1 | B11 100.000 | B9 100.000 | B9 100.000 | Prcp 100.0000 |
| 2 | B12 64.399 | B11 42.885 | Tair 59.627 | Tair 97.2280 |
| 3 | B3 43.114 | B6 40.049 | B6 50.805 | B11 5.2040 |
| 4 | B4 37.466 | B12 29.381 | B11 36.805 | B12 2.8079 |
| 4 | B2 31.412 | B1 28.338 | B12 25.088 | B2 1.8420 |
| 5 | B5 23.261 | B8A 24.860 | B7 20.342 | B8 1.3367 |
| 6 | B1 17.435 | B8 23.439 | B8 19.528 | B9 1.2699 |
| 7 | B9 14.423 | B7 18.883 | Prcp 19.268 | B4 1.1931 |
| 8 | B8 10.982 | B2 12.494 | B1 18.114 | B3 0.6025 |
| 9 | B6 7.127 | B5 5.806 | B8A 15.191 | B1 0.5769 |
| 10 | B7 1.218 | B3 1.983 | B2 6.849 | B8A 0.3501 |
| 11 | B8A 0.000 | B4 0.000 | B5 2.416 | B5 0.3275 |
| 12 | | | B3 1.751 | B6 0.0633 |
| 13 | | | B4 0.000 | B7 0.0000 |

## 4. CONCLUSIONS

Our study shows the use of plot scale sap flow measurements for modelling forest transpiration. For up-scaling transpiration from plot to landscape, it is essential to consider local meteorological situations, especially in mountainous regions where the topography strongly influences the weather. Additionally, being remote sensing data quality potentially affected by topography, specific corrections should be tested. Concluding, our results demonstrate the relevance of integrating *in-situ* data and remote sensing at compatible spatial scales for better understanding data-driven upscaling of ecosystem water fluxes. Such, if applied to a broader network of sites and climatic conditions, such an approach could offer unprecedented possibilities for investigating regional resilience and resistance capacity of our forests to an intensified hydrological cycle in the contest of a changing climate.